# Incipient charge order observed by NMR in the normal state of YBa$_2$Cu$_3$O$_y$

Tao Wu[1], Hadrien Mayaffre[1], Steffen Krämer[1], Mladen Horvatić[1], Claude Berthier[1], W.N. Hardy[2,3], Ruixing Liang[2,3], D.A. Bonn[2,3], and Marc-Henri Julien[1]

*1 Laboratoire National des Champs Magnétiques Intenses, UPR 3228, CNRS-UJF-UPS-INSA, 38042 Grenoble, France*

*2 Department of Physics and Astronomy, University of British Columbia, Vancouver, BC V6T 1Z1, Canada correlated*

*3 Canadian Institute for Advanced Research, Toronto, Ontario M5G 1Z8, Canada*

**The pseudogap regime of high-temperature cuprates harbours diverse manifestations of electronic ordering whose exact nature and universality remain debated. Here, we show that the short-ranged charge order recently reported in the normal state of YBa$_2$Cu$_3$O$_y$ corresponds to a truly static modulation of the charge density. We also show that this modulation impacts on most electronic properties, that it appears jointly with intra-unit-cell nematic, but not magnetic, order, and that it exhibits differences with the charge-density-wave observed at lower temperatures in high magnetic fields. These observations prove mostly universal, they place new constraints on the origin of the charge-density-wave and they reveal that the charge modulation is pinned by native defects. Similarities with results in layered metals such as NbSe$_2$, in which defects nucleate halos of incipient charge-density-wave at temperatures above the ordering transition, raise**

**the possibility that order-parameter fluctuations, but no static order, would be observed in the normal state of most cuprates if disorder were absent.**

**INTRODUCTION**

Unconventional superconductivity arises more often than not in the vicinity of another ordered electronic state. Over the last decade, a wealth of experimental results (1-11) has suggested the presence of ordered states in the region of the cuprate phase diagram in which a pseudogap is observed ("pseudogap regime"). Yet, the importance of these observations has remained controversial.

First, the breaking of different (translational, point-group or time-reversal) symmetries at temperatures not always coinciding with one another indicates a complicated sequence of phenomena whose mutual connections and microscopic origin are still elusive. Second, the relationship between the spectroscopic pseudogap and these putative orders has been much debated. Is the pseudogap produced by an ordered state (or by fluctuations thereof) or are they both consequences of the same physics? Third, some of the results obtained in given cuprate families have yet to be confirmed in other families or, conversely, some of the results obtained with a particular probe have not been verified by other techniques. This has raised legitimate doubts on the intrinsic and universal nature of ordering phenomena.

Tremendous progress in recent years is giving a new twist to the field. In particular, charge order competing with superconductivity has been discovered in $YBa_2Cu_3O_y$, using nuclear magnetic resonance (NMR) in strong magnetic fields $H\|c$ that significantly reduce the superconducting transition temperature $T_c$ (12,13). Because oxygen-ordered $YBa_2Cu_3O_y$ has particularly weak disorder, this result has been argued (12) to support the, hitherto controversial, view that underdoped cuprates are generically unstable towards charge ordering (14). Since the charge ordering transition temperature $T_{\text{charge}}$ can be as high as the superconducting transition $T_c$ in zero-field (12,13,15), the observed order clearly emanates from charge-density-wave (CDW) correlations developing in the normal state. Subsequent X-ray diffraction (XRD) experiments (16-20) have indeed found (incommensurate and short-ranged) CDW correlations far above $T_c$ and they have confirmed their competition with superconductivity.

These results raise critical questions for understanding electronic properties in the pseudogap regime and the superconductivity that emerges from it. What is the connection between the CDW modulation at low temperatures/high fields and that in the normal state and/or at low fields? Does the pseudogap regime host dynamic CDW correlations (a "fluctuating CDW order") or a thermodynamic CDW phase? How does this CDW relate to the nematic and magnetic orders claimed to occur near or above the CDW onset (2,3,10,21)?

Here, we report the simultaneous probe of these three (CDW, nematic and magnetic) orders using a single experimental technique, NMR, in $YBa_2Cu_3O_y$. We find evidence of a charge-density modulation that, although distinct from the CDW observed at lower temperatures in high fields, is already static in the normal state. The planar anisotropy of this modulation produces an intra-unit-cell imbalance between oxygen sites, which can be interpreted as a form of nematic order accompanying the short-range CDW order. This static charge modulation is argued to be universally present in the pseudogap regime of the cuprates where it is responsible for many anomalies in their electronic properties. In contrast, static magnetic order proves to be invisible at the NMR timescale. Finally, we find that the charge modulation is pinned by native defects, even in this extremely clean cuprate. This reveals the ubiquitous role of quenched disorder in shaping the experimental manifestations of electronic ordering. In particular, our results raise the possibility that the static charge modulation in the normal state only corresponds to 'frozen' CDW fluctuations, that is, an incipient CDW order that would be purely fluctuating in the absence of disorder.

## RESULTS

### The CDW modulation is static in the pseudogap state

The two studied crystals have ortho-II and ortho-VIII superstructures (Fig. 1) and hole-doping levels $p = 0.109$ and $0.125$, respectively (see methods). The central result of this work (Figs. 2,3) is the discovery of a quadrupole broadening $\delta\nu_{quad}$ of NMR lines, which is the signature of a spatial inhomogeneity of the charge distribution that grows on cooling below $T_{onset} \approx 140\text{-}170$ K (see methods). This effect has smaller magnitude and more gradual temperature dependence than the line splitting attributed to long-range charge order below $T_{charge} \leq 50\text{-}70$ K in high magnetic fields (12,13). That the increase of $\delta\nu_{quad}$ has a similar $T_{onset}$ and temperature dependence as the XRD intensity (Fig. 4) shows that the NMR line is broadened by the CDW-type modulation and not by nanoscale inhomogeneity of the hole-doping. As expected above $T_c$ and in agreement with XRD, this modulation is field-independent so that it should be present in zero-field (Fig. 3e).

That CDW-related effects previously observed by XRD are also detected, for the first time, by NMR in the normal state of a particularly clean high-$T_c$ cuprate is significant because it demonstrates that there is a charge modulation that is truly static below $T_{onset}$. The timescale set by the typical width of the $^{17}$O lines is indeed as low as ~10 kHz.

The static nature of the charge modulation has several consequences. First, this is pivotal for understanding its impact on many different properties (see later). Second, this sheds new light on the CDW phase diagram: since NMR sees the same $T_{onset}$ as XRD, there cannot be a timescale issue that would make $T_{charge} \approx 60$ K appear as high as $T_{onset} \approx 150$ K at the much shorter timescale of XRD. Therefore, $T_{charge}$ and $T_{onset}$ are two distinct temperature scales in the field *vs.* temperature phase diagram (Fig. 5). Third, the occurrence of a Kerr rotation (3) at the same $T_{onset}$ as the static NMR broadening (Fig. 4b) shows that the coincidence of Kerr and XRD onsets was not accidental. This supports the idea (22) that the breaking of certain symmetries by the charge pattern is required to produce the, yet unexplained, Kerr rotation. Fourth, neither of the following standard situations seems to be realized here: short-ranged CDW fluctuations above a CDW transition at $T_{CDW}$ (sometimes called 'fluctuating order') or static and long-range CDW order below $T_{CDW}$. That the charge modulation is static but short-ranged (16-20) necessarily implies that quenched disorder plays a role, even in such a clean cuprate.

**Microscopic nature of the CDW modulation**

Our data also brings new information on the structure of the CDW.

First, the charge modulation of the high-field phase develops on top of the normal-state charge modulation so that $T_{charge}$ does not correspond to a simple switch between two different CDW patterns. Indeed, the normal-state broadening actually persists below $T_{charge}$, where it is superimposed on the line splitting and is approximately temperature-independent (Fig. 3d).

Second, the modulation in the normal state is not just a weak-amplitude version of the modulation in the high-field phase. At least for the ortho-II sample, it shows less in-plane anisotropy than in the high-field phase where planar anisotropy in the magnitude and/or the period of the charge order is inferred from the difference in the splitting of Cu(2E) and Cu(2F) lines (13). In contrast, these two sites experience similar broadening here in the normal state (Fig. 3f).

Third, the charge modulation in the normal state breaks intra-unit-cell symmetry. Indeed, the raw data (Fig. 2) show that the O(2) sites (in bonds oriented along the *a* axis) experience a different broadening as compared to the O(3) sites (along *b*). At high temperatures, the difference of width between O(2) and O(3) is small and temperature-independent (Figs. 2a,b & 6), consistent with a differentiation due to the lattice anisotropy of YBa$_2$Cu$_3$O$_y$ (orthorhombicity and/or to strain from the chains). However, the width difference becomes temperature dependent and increases on cooling (Figs. 3a,b & 6), which shows that it involves electronic correlations. In this sense, the temperature-dependent differentiation represents a form of nematic order, even though it could eventually arise in response to the structural anisotropy.

**Origin of intra-unit-cell nematic order**

An important observation for understanding the microscopic origin of the inequivalence of O(2) and O(3) is that it appears at $T_{\text{onset}}$ (Figs. 3a,b & 6). The intra-unit-cell inequivalence thus appears as a mere consequence of the CDW pattern.

Two main types of patterns can produce such nematicity: a uniaxial CDW and a biaxial CDW with *d*-symmetry (23,24). In the latter case, however, the distribution of charge density is statistically identical for the two subsets of oxygen sites in orthogonal bonds. Therefore, the purely biaxial *d*-CDW cannot be visible in the NMR linewidth which represents the breadth of this statistical distribution. We conclude that the inequivalence of O(2) and O(3) in the NMR data of ortho-VIII (Fig. 6c,d) betrays a certain degree of planar anisotropy of the local charge modulation which has not been detected in the XRD measurements (18). Therefore ortho-VIII may not be fundamentally dissimilar to ortho-II for which a marked anisotropy is evident from both NMR (Figs. 3a,b & 6a,b) and XRD (19,20).

As described above, because it gives access to the width of a statistical distribution and not to a direct visualisation of each unit-cell, NMR may not be sensitive to all types of differentiations. Thus, the oxygen differentiation in the NMR data of $YBa_2Cu_3O_y$ might be, at least partially, different from the (long-ranged) intra-unit-cell nematicity identified by scanning-tunneling-microscopy in Bi-2212 (10). Also, we cannot exclude that nematic order exists without concomitant translational symmetry breaking (25) elsewhere in the phase diagram. Therefore, that the intra-unit-cell inequivalence measured here is nothing else than planar anisotropy of the short-ranged charge modulation is not necessarily a conclusion that can be claimed to be generic from our results.

**Lessons from canonical CDW systems**

In layered CDW materials such as $NbSe_2$, a generalized version of the Friedel oscillations of common metals takes the form of a static charge-density modulation of period $\pi/q_{CDW}$, where $q_{CDW}$ is the ordering wave-vector of the CDW, even above the CDW transition at $T_{CDW}$ (26-28). On cooling towards $T_{CDW}$, the charge modulation grows around defects over a typical length set by the charge correlation length $\xi_{charge}$ of the pure system and with an intensity reflecting the amplitude of the CDW susceptibility $\chi_{CDW}$ (26-28). This produces an NMR line broadening starting to be detected near $T_{onset}$ $\approx 2.3\ T_{CDW}$ and actually persisting below $T_{CDW}$ as the CDW pattern remains perturbed around defects (26-27).

All aspects of the above-described NMR broadening in YBa$_2$Cu$_3$O$_y$, including its onset near $T_{onset} \approx$ 2-3$T_{charge}$, strikingly resemble the above description. This suggests that the static charge modulation in the normal state also arises from CDW fluctuations frozen or "pinned" by defects which are always present in any sample. Furthermore, disorder-pinning offers a plausible explanation of the stronger in-plane anisotropy of the modulation in the high-field phase of ortho-II: an intrinsically-unidirectional modulation in a square lattice becomes locally bidirectional when pinned to defects (14,29,30); we expect this pinning to affect more strongly the short-range order of the normal state than the long-range order of the high-field phase.

**Generic impact of the static charge modulation**

Because the pinned modulation is static and pervasive in CuO$_2$ planes, it offers a natural explanation of the anomalous $T_c$ depression near $p$ = 0.12 and of many anomalies observed near $T_{onset}$ in YBa$_2$Cu$_3$O$_y$: the Hall coefficient curves downward, suggesting the beginnings of a Fermi surface reconstruction (11), the electrical resistivity $\rho_{ab}(T)$ shows an inflexion point (31), new modes appear in optical spectroscopy (32,33). We also note that the NMR relaxation rate $(T_1T)^{-1}$ has a maximum near $T_{onset}$, until now attributed to the pseudogap (ref. 34).

There is moreover, both direct and indirect, evidence that similar disorder-pinned charge modulation is ubiquitous in the normal state of underdoped cuprates: similar Fermi surface reconstruction suggesting incipient charge ordering (35) and similar sequence of temperature crossovers in $\rho_{ab}$ (36) are observed in HgBa$_2$CuO$_{4+\delta}$. Out-of-plane oxygen defects are thought to play a primary role in pinning analogous short-ranged charge modulations in Bi-based cuprates (37). Likewise, particle-hole asymmetry and spectral broadening in photoemission experiments have been attributed to short-range, rather than long-range, order whose microscopic nature could not be identified (4).

**Charge-density-wave *versus* pseudogap**

In general, caution is required when discussing onset temperatures as they can be influenced by the signal-to-noise ratio in the data and by the subjectivity of the chosen criterion. Here, we cannot exclude that $T_{onset}$ values would be higher in measurements with orders of magnitude greater precision. Also, the pseudogap onset temperature is not unequivocally defined in $YBa_2Cu_3O_y$: for the same y value, $T^*$ values can differ by ~100 K depending on the measurement (2,31,38). Nonetheless, even with these precautions, the experimental situation is unambiguous: $T_{onset}$ and $T^*$ appear as two dissimilar, markedly separated, scales in the phase diagram (Fig. 7), especially since $T^*$ defined by the onset of a decrease in the spin susceptibility, is definitely above room temperature for our $YBa_2Cu_3O_y$ samples (38). It should thus be clear that there is nothing in our data suggesting that CDW correlations could be responsible for the pseudogap.

This, however, does not mean that the two phenomena are entirely unrelated. Since CDW correlations are manifested only within the pseudogap regime, they must have roots in common (likely local antiferromagnetic correlations). Furthermore, we cannot exclude that CDW correlations contribute to the pseudogap in the temperature range $T_c < T < T_{onset}$, even though they are not involved in its formation at higher temperature. The question of the relationship between the pseudogap and the CDW is actually broader and more complex than a simple issue of causation. It is thus beyond the scope of this paper.

**Search for intra-unit-cell magnetic order**

Intra-unit-cell (*i.e.* $Q = 0$) magnetic order is observed in neutron scattering studies (2) and its onset turns out to coincide with thermodynamic evidence of a phase transition at $T^*$ (21). This could lead to the conclusion that $Q = 0$ magnetic order is a genuine broken symmetry, and the only such, associated with the pseudogap onset. However, previous attempts at detecting the corresponding magnetic fields with magnetic resonance techniques proved unsuccessful (39-41).

In order to test the occurrence of magnetic order, our work presents advantages never met before in a single study. First, NMR, unlike muon-spin-rotation, cannot be suspected to electrically perturb $CuO_2$ planes. Second, $YBa_2Cu_3O_y$, unlike $YBa_2Cu_4O_8$, has proven magnetic scattering at $Q = 0$ (2). Third, unlike in $HgBa_2CuO_{4+\delta}$, magnetic fields in $YBa_2Cu_3O_y$ should be induced not only at apical oxygen sites but also at planar sites (42). Fourth, our ortho-II crystal is virtually identical to the crystal showing a thermodynamic transition (21).

In our ortho-II sample, the temperature and field dependence of the line width (Fig. 8) implies that any static field at $T = 60$ K cannot be larger than ~0.3 G at apical O(4E) sites (as in Hg1201 (ref. 41)) and ~4 G at the planar O(2) sites. Such upper bounds are about two orders of magnitude smaller than a recent theoretical estimate for static current-loop order (42).

How to reconcile the absence of magnetic order in NMR with the neutron scattering data? A possibility is that the internal field fluctuates slow enough to appear static in neutron scattering (2) but too fast to impact NMR spectra. It has been proposed that this happens if defects limit the correlation length (43). In such event, we expect a static field to subsist but its reduced amplitude, as compared to the clean case, may be too small to be detected. Again, disorder may be a key-aspect of the problem.

**DISCUSSION**

Let us first briefly summarize our main findings. At doping levels $p$ = 0.109-0.125, the spatial modulation of the charge-density in the normal state of YBa$_2$Cu$_3$O$_y$ is static, albeit coherent over a finite length only (16-20), and pinned by disorder. It includes a form of intra-unit-cell nematic order as its planar anisotropy leads oxygen sites in perpendicular bonds to experience statistically different amplitudes of the charge modulation. The modulation is distinct from the long-range CDW order that develops on top of it at low temperatures and high-fields.

The observed charge modulation could arise from static and long-range CDW order made short-ranged by disorder (44). This possibility may be supported by the much longer CDW correlation length $\xi_{charge} \approx 16a$ at $T \approx 2T_{charge}$ in the normal state of YBa$_2$Cu$_3$O$_y$ (18) as compared to $\xi_{charge} \approx 3a$ at $T \approx 2T_{CDW}$ in NbSe$_2$ (29). This observation and recent XRD work (45) appear consistent with a domain structure, as expected from weak, collective, pinning of an ordered CDW (46). Weak pinning may itself be expected in YBa$_2$Cu$_3$O$_y$ from its well-known chain-oxygen defects. Such out-of-plane defects in the texture of the charge reservoir can impact the electronic texture of CuO$_2$ planes and they are relatively dense: even the best-ordered ortho-II samples contain oxygen defects at the % level. In this hypothesis, $T_{onset} \approx 160$ K is the vestige of the sharp CDW phase transition that would be observed in the absence of disorder whereas $T_{charge}$ is a secondary CDW transition, such as a commensurate lock-in (44).

On the other hand, the above-discussed striking analogy with NbSe$_2$ is suggestive of an opposite explanation: CDW fluctuations become frozen by disorder on approaching a putative CDW transition at $T_{CDW} \ll T_{onset}$. The transition observed at $T_{charge} \approx 60$ K could then naturally correspond to $T_{CDW}$ but it is observed only in high magnetic fields because superconductivity otherwise impedes the growth of CDW correlations in $YBa_2Cu_3O_y$. The largish $\xi_{charge}$ would indicate a particularly broad pre-transitional regime of CDW fluctuations in $YBa_2Cu_3O_y$, for whatever reason (46,47). In this scenario, no genuine (that is, long-range) charge order would occur in the (pristine) normal state. Nevertheless, there would be (fluctuating) CDW correlations as these are undoubtedly intrinsic and not due to disorder.

Ultimately, we find no conclusive evidence of a static, long-range, electronic order that would be universally present in the normal state of cuprates. Whether charge order, or any other order, would be purely fluctuating or long-ranged in the absence of disorder emerges as a central question for understanding the pseudogap regime and for discriminating theoretical proposals (23-25,44,48-54), but how to decide? An intrinsic problem with the cuprates is that off-stoichiometry makes any clean-limit out-of-reach so that varying the amount of disorder may not bring unequivocal answers. Although ortho-II $YBa_2Cu_3O_{6.56}$ may be regarded as quasi-stoichiometric (and actually shows twice less quadrupole broadening between $T_{onset}$ and $T_c$ than the more disordered ortho-VIII – see Fig. 3b,c), a solution to the conundrum might come from the even more stoichiometric $YBa_2Cu_4O_8$, for which there is also robust evidence of CDW correlations (11,55). In every case, by placing the role of disorder under the spotlight and by identifying remaining key-questions our work offers a roadmap for elucidating the enigma of electronic-ordering phenomena in the pseudogap regime of the cuprates.

*Note added in proof*: during the review process of this manuscript, new X-ray experiments have been published, with the following conclusions: in $YBa_3Cu_3O_y$, the spatial extent of the short-range CDW correlations is insensitive to the level of quenched oxygen disorder for ortho-VIII and ortho-III (57) and the hole-doping dependence of $T_{onset}$ has a dome-like shape around $p = 0.12$ (58,59) suggesting that the Kerr onset precedes the CDW onset below $p \approx 0.11$ (see Fig. 7). This latter result raises the question as to whether the breaking of point-group and translational symmetries occurs at two distinct temperatures below $p \approx 0.11$, unlike what we observe here for $p \approx 0.11$-0.125. Short-range CDW order is confirmed to occur in the bulk of Bi2201 (60), Bi2212 (61,62) and Hg1201 (63).

## METHODS

### Samples

High quality, oxygen-ordered, detwinned single crystals of YBa$_2$Cu$_3$O$_y$ were grown in non-reactive BaZrO$_3$ crucibles from high-purity starting materials; see refs. (3,10-13,15,17-19,21,32,33,39) for works on similar crystals. The two crystals studied here were enriched with the oxygen-17 ($^{17}$O) isotope which has a nuclear spin $I = 5/2$ (because $I = 0$ of $^{16}$O is not observable by NMR). High field NMR experiments have been performed in these samples (13). YBa$_2$Cu$_3$O$_{6.56}$ (ortho-II structure) has a hole-doping $p = 0.109$ and YBa$_2$Cu$_3$O$_{6.68}$ (ortho-VIII structure) has a hole-doping $p = 0.125$.

### NMR

Standard spin-echo techniques were used with a laboratory-built heterodyne spectrometer. Spectra were obtained at fixed magnetic fields in a superconducting magnet, by adding Fourier transforms of the spin-echo signal recorded for regularly spaced frequency values.

### Line width analysis

Electric field gradients created by on-site and surrounding charges interact with oxygen ($^{17}$O) and copper ($^{63}$Cu) nuclei through the electric quadrupole interaction. This produces "satellite" lines in their NMR spectra at relative positions set by the quadrupole frequency $\nu_{quad}$ (supplementary Fig. 1). The satellite width arises from a distribution of values, in the bulk of the sample, of the magnetic hyperfine field $\delta\nu_{magn}$ and of the quadrupole frequency $\delta\nu_{quad}$. Both $\delta\nu_{quad}$, which is related to the electric-field-gradient, and $\delta\nu_{magn}$ grow as the charge density becomes spatially inhomogeneous. For quantitative analysis, it is useful to disentangle them even though the effects described below are discernible in the raw data (Fig. 2).

The width $\delta\nu_{total}$ of each line reflects spatial distributions of quadrupole frequencies $\nu_{quad}$ and of hyperfine magnetic shifts $K = (\nu_{magn}-\gamma H_0)/\gamma H_0$ where $\gamma$ is the nucleus' gyromagnetic ratio and $H_0$ the external magnetic field. That the values of the spin-spin relaxation time $T_2$ are considerably larger than the inverse of the line width, and that the broadening is not identical for high and low frequency satellites, excludes broadening by a dynamical effect.

In the field-induced charge ordered state of $YBa_2Cu_3O_y$, local differences in the charge density produce changes in both $\nu_{quad}$ and $\nu_{magn}$ with respect to the homogeneous situation. We have analysed these changes in terms of splitting of the $^{63}Cu(2F)$ line, that is, the appearance of bimodal distribution of sites having a charge density higher or lower than average (12,13). The higher (lower) charge density increases (decreases) the values of both $\nu_{quad}$ and $\nu_{magn}$. This correlation produces a splitting for the high-frequency $^{63}Cu(2F)$ satellite that is larger than for the low-frequency satellite. Actually, for fields close to ~30 T, it turns out that the amplitude of the magnetic and quadrupole changes are nearly equal: $\Delta\nu_{quad} \approx \Delta\nu_{magn}$. This results in a large splitting of the high frequency $^{63}Cu(2F)$ satellite $\Delta\nu_{total} \approx \Delta\nu_{magn}+\Delta\nu_{quad} \approx 2\nu_{quad}$ and a vanishingly small splitting for the low frequency $^{63}Cu(2F)$ satellite $\Delta\nu_{total} \approx \Delta\nu_{magn}-\Delta\nu_{quad} \approx 0$. A similar effect is observed with $^{17}O$ NMR: no splitting induced by charge-order can be resolved for the low frequency satellites in the investigated field range.

As Fig. 2 shows, the contrast in the broadening between high and low frequency satellites in the normal state is reminiscent of the contrast in the splitting observed in the field-induced charge-ordered state. This is direct evidence that the broadening reflects a distribution of $\nu_{quad}$ and $\nu_{magn}$ values which are correlated as in the charge-ordered state: nuclei with values of $\nu_{quad}$ larger (smaller) than average also have a larger (smaller) $\nu_{magn}$ than average. A similar observation was previously made in $La_{2-x}Sr_xCuO_4$ (56) but the considerably higher resolution of our NMR spectra in $YBa_2Cu_3O_y$ single crystals allows quantitative analysis of the data and deeper physical insight.

If these two distributions were uncorrelated, we would express the line width $\delta\nu$ with an ad-hoc formula, appropriate for $YBa_2Cu_3O_y$:

$$(\delta\nu_{total})^2 = (n\,\delta\nu_{quad})^2 + (\delta\nu_{magn})^2 \qquad (1)$$

with for $n = 1$ (2) for the first (second) satellites and $n = 0$ (in reality, it is $\delta\nu_{quad} \approx 0$) for the central line of both planar $^{17}O$ and $^{63}Cu$ sites (provided $H\|c$ for the latter, as is the case for the data here). Here, in contrast, we assume the simplest, *i.e.* linear, correlation between the two contributions and thus now rewrite the width of the $^{17}O$ or $^{63}Cu$ NMR lines as:

$(\delta\nu_{total})^2 = (\delta\nu_{magn})^2 + (\delta\nu_{magn0})^2$ for each central line,

$(\delta\nu_{total})^2 = (\delta\nu_{quad} + \delta\nu_{magn})^2 + (\delta\nu_{magn0})^2$ for each first high frequency satellite,

$(\delta\nu_{total})^2 = (2\delta\nu_{quad} + \delta\nu_{magn})^2 + (\delta\nu_{magn0})^2$ for each second high frequency satellite,

$(\delta\nu_{total})^2 = (\delta\nu_{quad} - \delta\nu_{magn})^2 + (\delta\nu_{magn0})^2$ for each first low frequency satellite,

$(\delta\nu_{total})^2 = (2\delta\nu_{quad} - \delta\nu_{magn})^2 + (\delta\nu_{magn0})^2$ for each second low frequency satellite.

By doing so, we assume that, in addition to the two coupled terms ($\delta\nu_{quad}$ and $\delta\nu_{magn}$), there is an uncoupled, "natural", magnetic contribution to the broadening ($\delta\nu_{magn0}$), but no uncoupled quadrupole contribution. For the ortho-II sample, this assumption is justified by the experimental data of Fig. 2 where the similar linewidth values for the central line and for the second satellites at high temperature indeed demonstrate that most of the broadening near room temperature is magnetic. For the ortho-VIII sample, we maintain this assumption in order to limit the number of parameters, although quadrupole broadening at room temperature is not negligible.

With the above set of equations and with experimental data $\delta\nu_{\text{total}}$ for the different lines, the values of $\delta\nu_{\text{quad}}$, $\delta\nu_{\text{magn}}$ and $\delta\nu_{\text{magn0}}$ can be extracted as a function of temperature. For the ortho-II sample, the central line data can be used because O(2) and O(3) central lines are separated. For the ortho-VIII sample, on the other hand, the overlap between the central lines of different sites forces us to rely on the satellite data only, *i.e.* we have a set of four equations instead of five.

A, relatively small asymmetry between high and low satellite widths, and thus a coupling between quadrupole and magnetic broadenings, already exists at room temperature, in both ortho-II and ortho-VIII for O(3E) and/or O(3F) sites (Fig. 2), and it is slightly temperature dependent in ortho-VIII. This is because trivial inhomogeneity of the hole-density also leads to coupled magnetic and quadrupole broadenings, reflecting the spatial distribution of spin and charge densities, even if these modulations occur on a macroscopic scale and are not necessarily periodic. In contrast, the phenomenon that we analyse in this paper occurs below a well-defined crossover (clearly visible in the range 140-170 K in the raw data) and the corresponding $T_{\text{onset}}$ values match the onset temperatures for CDW correlations observed by X-ray scattering at two different doping levels.

Despite its shortcomings, this analysis correctly captures the salient features of the quadrupole broadening in YBa$_2$Cu$_3$O$_y$, namely the marked increase of $\delta\nu_{\text{quad}}$ between $T_{\text{onset}}$ and the superconducting transition $T_{\text{c}}$. Furthermore, this simple modelling remarkably accounts for the very atypical magnetic field-dependence of the width of the different lines (supplementary Fig. 2). On general grounds, the field dependence results from the balance between quadrupole and magnetic contributions to the broadening, which are field-independent and linearly dependent on field, respectively. Here again, the striking contrast between high and low frequency satellites as a function of field is direct evidence for the presence of a quadrupole broadening which is coupled to a magnetic hyperfine broadening (supplementary Fig. 3 for the ortho-VIII sample). Finally, the validity of our analysis is further supported by very consistent results obtained for three different field orientations (Fig. 3a,b) as well as for different field strengths (Fig. 3e and supplementary Fig. 4). It should be noted that the precise determination of $\delta\nu_{quad}$ is important for comparison purposes (O(2) *vs*. O(3) sites, ortho-II *vs*. ortho-VIII) but not for the main conclusions of our paper. These essentially rely on the existence of a well-identified temperature crossover in the quadrupole broadening, which is seen clearly already in the raw data.

**Evidence of intra-unit-cell nematic order**

Unlike in ortho-II, the above three-parameter analysis in ortho-VIII leads to unphysical saturation of $\delta\nu_{quad}$ values for the O(2EF) site at temperatures 70 K < $T$ < 90 K. While this does not affect the determination of $T_{onset}$, it prevents an accurate comparison with the results for O(3F). Thus, in order to compare the broadening of the two sites, we use a cruder two-parameter analysis, simply writing $\delta\nu_{total} = n\delta\nu_{quad} \pm \delta\nu_{magn}$ with $n = 1$ (2) for the first (second) satellites and the sign plus (minus) applies to the high (low) frequency satellites. Above 90 K, the $\delta\nu_{quad}$ values obtained with this simplified analysis turn out to be close to those obtained with the three-parameter analysis.

For both ortho-II and ortho-VIII, O(3) high and low frequency satellites have slightly different width above $T_{onset}$, while O(2) pairs of satellites have the same width. Such contrast between O(2) and O(3) above $T_{onset}$ is likely of structural origin. Below $T_{onset}$, the magnetic contribution to the broadening causes a large width difference between the high and low frequency satellites of O(2) sites. For O(3) sites on the other hand, this difference is much smaller in ortho-VIII and it is strikingly absent in ortho-II (supplementary Fig. 5). This indicates that short-range CDW order affects much less the spin susceptibility at O(3) than at O(2) sites. See complementary data in supplementary Figs. 2,3.

**Determination of an upper bound on static magnetic fields**

Of our full set of data for O(2) and O(3) sites in three different field orientations in the ortho-II sample, the data shown in Fig. 8 are those showing the maximum broadening (~10 kHz, $H \| a$) between $T_{onset}$ and 60 K. We thus analyse this set of data. The linear field-dependence of $\delta\nu_{central}$ indicates that most of this broadening arises from inhomogeneity in the electronic spin polarization. $\delta\nu_{central}$ can be separated into two terms: $\delta\nu_{central}(H) = [(\delta\nu_{spin}(H))^2 + (\delta\nu_{extra})^2]^{1/2}$ where $\delta\nu_{spin} \propto H$ is due to the paramagnetic response of electronic moments and $\delta\nu_{extra}$ is due to all field-independent contributions, including any additional field created by putative circulating currents or by antiferromagnetically-ordered spins at O sites. Lines correspond to fits to $\delta\nu_{central}(H) = [\alpha H^2 + (\delta\nu_{extra})^2]^{1/2}$. For $H \| a$, $\delta\nu_{extra} = 2.4 \pm 1.3$ kHz. Within error bars, this defines a maximum possible intercept $\delta\nu_{extra}(max) \approx 2.4 + 2*1.3 = 5$ kHz. After subtraction of the "homogeneous" line width $\delta\nu_{homo} \approx 1/T_2 \approx 1$ kHz, where $T_2$ is the nuclear spin-spin relaxation time, and division by a factor of 2 as the broadening must be caused by an unresolved line-splitting due to fields pointing in opposite directions, a maximum frequency shift of $\nu_{splitting}(max) = 1/2(\delta\nu_{extra}(max)^2 - \delta\nu_{homo}^2)^{0.5} \approx 2.4$ kHz is obtained. Using the $^{17}$O gyromagnetic ratio of 0.5772 kHz G$^{-1}$, $\nu_{\text{splitting}}$(max) translates into that a maximum static field of 4 G at the planar O sites.

(Date received……………………………….)

Acknowledgements:

Acknowledgments: We thank W. Atkinson, E. Dalla Torre, S. Fratini, M. Hirata, M. Hücker, A. Kampf, E.A. Kim, D. LeBoeuf, M. LeTacon, M. Mali, P. Monceau, J. Orenstein, C. Proust, and especially S. Kivelson, for useful exchanges. This work was funded by Université J. Fourier – Grenoble and by the French Agence Nationale de la Recherche (ANR) under reference AF-12-BS04-0012-01 (Superfield).

Authors' Contributions: W.N.H., R.L. and D.A.B. prepared the samples. T.W. performed the experiments with help from H.M. & M.H.J. and technical inputs from H.M., S.K. & M.H.. T.W. & M.H.J. analysed and interpreted the data. C.B. provided conceptual advice. M.H.J. wrote the paper and supervised the project. All authors discussed the results and commented on the manuscript.

The authors declare no competing financial interests. Correspondence and requests for materials should be addressed to M.-H.J. (marc-henri.julien@lncmi.cnrs.fr).

**Figure 1 | Sketch of ortho-II and ortho-VIII structures.**

**a**,**b**, inequivalent O and Cu sites. Cu(2F) and O(3F) lie below oxygen-full chains, Cu(2E) and O(3E) lie below empty chains. Grey rectangles represent the super-cell due to ortho-II and ortho-VIII oxygen order.

**Figure 2 | Qualitative evidence of static charge modulation.**

**a**,**b**,**c**,**d** Full-width-at-half-maximum (FWHM) of the central line and of high/low frequency quadrupole satellites in ortho-II ($H$ = 12 T for $^{17}$O, 15 T for $^{63}$Cu). O(3) stands for either O(3E) or O(3F) as they have identical width (see supplementary Fig. 7). **e**,**f**, FWHM data in ortho-VIII ($H$ = 15 T). The satellite data correspond to the (±5/2,±3/2) transitions for $^{17}$O ($H||b$) and to the (±3/2,±1/2) transitions for $^{63}$Cu ($H||c$). The grey bar marks the onset of broadening due to short-ranged CDW. Qualitative evidence for a static charge modulation lies in the different temperature dependence between the central line and any of the satellites and in the different width of high and low-frequency satellites (see methods). To help visualisation, the black trace corresponds to the central line data (stars) shifted so as to overlap with satellite data at high temperature. Qualitative evidence for intra-unit-cell nematicity comes from the different broadening for O(2) and O(3) sites.

**Figure 3 | Quantitative evidence of static charge modulation.**

**a-f,** Quadrupole contribution to the line broadening $\delta\nu_{\text{quad}}$ extracted from the width data of Fig. 2 (see methods for details on the analysis and supplementary Figs. 2,3,5 & 6 for complementary data). **a,b** $^{17}$O(2) data and $^{17}$O(3) data in ortho-II. Triangles and circles represent data for $H||b$ (multiplied by 0.78 for O(2)), $H||a$ (multiplied by 0.78 for O(3)) and $H||c+20°$ (multiplied by 1.5 for both sites), respectively. **c,** $^{17}$O(3F) data in ortho-VIII ($H||b$). Note the twice-larger change of $\delta\nu_{quad}/\nu_{quad}$ between $T_{onset}$ and $T_c$ as compared to ortho-II. **d,** red dots are the same O(2), $H||c+20°$, data as in **a** (over a smaller temperature range) and black squares are data for the split peaks in the high field phase $H = 28.5$ T. The broadening of the normal persists at low temperature in the high-field phase. **e**, O(2) data vs. field ($H||c+20°$) at T = 60 K. **f**, $^{63}$Cu data. Up (down) symbols correspond to $H||c$ = 15 T (11 T). The identical broadening for the Cu(2E) and Cu(2F) sites is consistent with the identical broadening of O(3E) and O(3F) (see supplementary Fig. 7). Error bars are s.d. in the fits of the lineshapes.

**Figure 4 | Connecting NMR with X-ray and Kerr-effect data.**

**a**,**b**, NMR quadrupole broadening (from Fig. 2a), X-ray scattering intensity (20) and Kerr angle (3) data scaled in ortho-II samples (slight doping differences explain the different $T_{onset}$ values). See supplementary Fig. 8 for similar scaling of the ortho-VIII data.

**Figure 5 | Field–temperature phase diagram of charge-ordered YBa$_2$Cu$_3$O$_y$.**

The high temperature/low field region is dominated by static but short-range CDW order. The question raised in this work is whether this should be explained by CDW fluctuations frozen by quenched disorder (in which case, $T_{charge}$ is the only CDW transition) or by static and long-range CDW order disrupted by disorder (in which case $T_{onset}$ represents the primary CDW phase transition and $T_{charge}$ a secondary CDW transition).

**Figure 6 | Intra-unit-cell electronic nematicity.**

**a**, CDW-induced magnetic hyperfine contribution to the line broadening in ortho-II at O(2) and O(3) planar sites. See methods for analysis details. **b**, Nematicity amplitude $N$ calculated as the difference of [$\delta\nu_{magn}$(T)-$\delta\nu_{magn}$(HT)] values between O(2) and O(3) sites, where $\delta\nu_{magn}$(HT) is the constant high temperature limit shown by orange and blue lines. **c**, CDW-induced quadrupole contribution to the broadening in ortho-VIII. **d**, Nematicity amplitude $N$ calculated as the difference of [$\delta\nu_{quad}$(T)-$\delta\nu_{quad}$(HT)] values between O(2EF) and O(3F) sites, where $\delta\nu_{quad}$(HT) is the high temperature limit shown by orange and blue lines. Since the measurement was performed with $H||b$, the [$\delta\nu_{quad}$(T)-$\delta\nu_{quad}$(HT)] data for O(3) were multiplied by a factor 0.78 as in ortho-II (Fig. 3). Grey vertical bars indicate $T_{onset}$, the onset of NMR broadening by static short-range CDW order. Orange and blue curves guide the eye. Error bars are s.d. in the fits of the lineshapes. Note that neither the magnitude of the differentiation observed in the electronic spin polarisation, which is particularly striking in ortho-II (**a**), nor the magnitude of the differentiation in the quadrupole frequency, which is opposite for ortho-II (Fig. 3**a**,**b**) and ortho-VIII (**c**), are understood. These quantitative data represent unique information on charge order at the microscopic level but exploiting them requires theoretical input on how the CDW affects the magnetic and electric fields measured in NMR.

**Figure 7 | Temperature-doping phase diagram of charge-ordered YBa$_2$Cu$_3$O$_y$.**

Phase diagram as a function of hole doping. The dashed line in the superconducting (SC) dome represents $T_{charge}$, the transition towards long-range CDW order, only observed under magnetic fields (15,16). $T^*$ represents the pseudogap onset. The question mark underlines the current uncertainty as to whether $T_{onset}$ decreases or increases with diminishing $p$ below $p \approx 0.11$ or whether it separates into two different scales according to the experimental probe. Error bars in the Kerr and X-ray data are from refs. 3,17,18 and 20. Error bars in the NMR data are defined by the error margin for scaling the NMR broadening to the Kerr data (Fig. 4 and Supplementary Fig. 8).

**Figure 8 | Search for intra-unit-cell magnetic order.**

**a**, FWHM of the apical $^{17}$O(4E) central line in ortho-II (data from this sample are used because of the sharper lines as compared to ortho-VIII). The absence of temperature dependence, to within ~0.4 kHz, provides an upper bound of 0.3 G for any additional magnetic field created at apical O sites at $T$ = 60 K. **b**, FWHM $\delta \nu_{central}$ of planar $^{17}$O(2) central line at $T$ = 60 K in ortho-II. These data define a maximum static field of 4 G at the planar O sites (see methods). Error bars are s.d. in the fits of the lineshapes